\documentclass[aps,showpacs,prd,twocolumn]{revtex4-2}
\usepackage{eurosym}
\usepackage{amsfonts}
\usepackage{amssymb}
\usepackage{amsmath}
\usepackage{color}
\usepackage[usenames,dvipsnames]{xcolor}
\usepackage{float}
\usepackage{accents}
\usepackage{graphicx}
\usepackage{epstopdf}
\usepackage[colorlinks=true,pdfstartview=FitV,linkcolor=blue,citecolor=blue,urlcolor=blue,breaklinks=true]{hyperref}

\begin{document}

\title{Self-dual solitons in a Born-Infeld baby Skyrme model}
\author{Rodolfo Casana}
\email{rodolfo.casana@ufma.br}
\email{rodolfo.casana@gmail.com}
\author{Andr\'{e} C. Santos}
\email{andre.cs@discente.ufma.br}
\email{andre$\_$cavs@hotmail.com}
\affiliation{Departamento de F\'{\i}sica, Universidade Federal do Maranh\~{a}o,
65080-805, S\~{a}o Lu\'{\i}s, Maranh\~{a}o, Brazil.}

\begin{abstract}

We show the existence of self-dual (topological) solitons in a gauged version of the baby Skyrme model in which the Born-Infeld term governs the gauge field dynamics. The successful implementation of the Bogomol'nyi-Prasad-Sommerfield formalism provides a lower bound for the energy and the respective self-dual equations whose solutions are the solitons saturating such a limit. The energy lower bound (Bogomol'nyi bound) is proportional to the topological charge of the Skyrme field and therefore quantized. In contrast, the total magnetic flux is a nonquantized quantity. Furthermore, the model supports three types of self-dual solitons profiles: the first describes compacton solitons, the second follows a Gaussian decay law, and the third portrays a power-law decay. Finally, we perform numerical solutions of the self-dual equations and depicted the soliton profiles for different values of the parameters controlling the nonlinearity of the model.
\end{abstract}

\maketitle

\section{Introduction}

The Skyrme model \cite{skyrme} is a well-known {low-energy field theory proposal to study} nonperturbative Quantum Chromodynamics (QCD). This nonlinear {effective field theory, defined initially} in $(3+1)$ dimensions, provides {some insights into the} physical properties of hadrons and nuclei {belonging to the realm of the low-energy QCD} \cite{adkins}. In this framework, the physical structures emerge as topological solitons so-called Skyrmions.

Over the years have seen remarkable progress in our understanding of Skyrme-like models, especially about its corresponding planar version, the {also known as} baby Skyrme model \cite{Piette}. The $(2+1)$-dimensional {version serves as a laboratory to study} many aspects of the original Skyrme model. {Besides, the skyrmions have also attracted the community's attention because they are used or come up in the description of various physical systems. For example,} in magnetic materials \cite{muhlbauer}, topological quantum Hall effect \cite{Soundhi}, chiral nematic liquid crystals \cite{Fukuda}, superconductors \cite{zyuzin}, brane cosmology \cite{Kodama}.

{The Skyrme model consists of a} $O(3)$ nonlinear sigma-model term (a quadratic kinetic term),  the Skyrme term (a quartic kinetic term), and a potential (a non-derivative one). For the baby Skyrme model, the presence of such a potential is obligatory to stabilize the soliton solutions \cite{Leese,Derrick}, being optional in the $3+1$-dimensional version. Although the standard baby Skyrme model describes stable solitons, it does not possess a self-dual or Bogomol'nyi-Prasad-Sommerfield (BPS) {structure. Nevertheless, lacking the sigma model term, the so-called restricted baby Skyrme model   \cite{Gisiger} does admit BPS configurations} \cite{adam}.

{To investigate the electric and magnetic properties of the Skyrme model, we must couple it with a $U(1)$ gauge field \cite{Witten}. Solitonic} solutions in the gauged baby Skyrme model {find firstly in \cite{gladikowski}, being later considered in \cite{Shnir} and, by including Chern-Simons term} in \cite{Shnir2,Lerida}. {On the other hand, in \cite{adam2,adam4,adam5} were studied the BPS configurations (carrying magnetic flux alone) in the restricted gauge baby Skyrme model. Moreover,  in \cite{adam6, Casana01, Casana02} is analyzed the self-dual solutions carrying {both the magnetic flux and electric charge.} It is essential to} point out the restricted gauged baby Skyrme model also allows a supersymmetric extension \cite{Bolognesi0, Bolognesi02, Queiruga00, Sasaki00, Queiruga}.

{In literature, the arising of topological defects also is a current issue in generalized or effective field theories. Principally, the new solitons may present quite different features than those found in the usual models. There are} many results in models characterized by possessing nonstandard kinetic terms, for example, in \cite{Babichev,DBazeia01,Babichev2,DBazeia02,RCasana1,RCasana3, Kuniyasu,CAdam,CAdam2}, including supersymmetric extensions \cite{CAdam3, Koehn}. {Moreover, we list additional applications, for example,} in the inflationary phase of the universe \cite{Armendariz1}, in strong gravitational waves \cite{Mukhanov}, tachyon matter \cite{ASen}, dark matter \cite{Armendariz2}, black holes \cite{Bronnikov, Faraoni}, and other topics \cite{Garriga, Scherrer}.

Among several generalized or effective models, we may highlight the Born-Infeld (BI) electrodynamics \cite{BornInfeld}, also dubbed Dirac-Born-Infeld theory \cite{Dirac}. In this elegant model, the gauge field's kinetic term is a highly nonlinear function of the Maxwell term (instead of the usual Maxwell term) that removes the divergence of the electron self-energy appearing in classical electrodynamics. The Born-Infeld-Higgs (BIH) model provides one of the pioneer
studies about vortex solutions within the scope of such generalized theories \cite{Shiraishi}. Later, in \cite{Moreno} was studied the existence of electrically charged BIH vortices through the inclusion of the Chern-Simons term. {Further, the BI theory has been explored} from different contexts, including topological defects \cite{DBazeia02, RCasana1, RCasana3, Kuniyasu, Dion, Pavlovskii}, unusual properties under wave propagation \cite{Boillat}, in gravitation and cosmology \cite{Garcia, Dey}, in quantum gravity \cite{Ellis}, and also supersymmetric extensions \cite{Queiruga}. Currently, the BI electrodynamics worthy of special attention since it appears in the low-energy limit of string/D-brane physics \cite{Fradkin, Brecher, Tseytlin, Sarangi, Babichev2}.

Thus, motivated by the versatility of the Born-Infeld term, the present manuscript aims to show the existence of the topological BPS solitons in a Born-Infeld baby Skyrme model and explore their main physical properties. We have organized the paper as follows. In Sec. II, we introduce the BI restricted baby Skyrme model and
implement the BPS formalism to get the topological energy lower-bound and self-dual equations that provide the soliton solutions saturating such a limit. In Sec. III, we focus our attention on configurations rotationally symmetric. We studied the behavior of the profiles around boundaries, performed the numerical solution of the BPS equations, and depicted the main features of the BPS configurations. Finally, in Sec. IV, we include our remarks and conclusions.

\section{The Born-Infeld restricted baby Skyrme model\label{themodel1}}

We consider a model describing the interaction between the restricted baby Skyrme field and a Born-Infeld gauge field defined by the Lagrangian
\begin{equation}
L=E_{0}\int d^{2}\mathbf{x}\,{\mathcal{L}},
\end{equation}%
being $E_{0}$ a common factor of energy scale which hereafter we set to be $E_{0}=1$, and $\mathcal{L}$ stands for the Lagrangian density given by
\begin{equation}
\mathcal{L}=\beta ^{2}\left( 1-\mathcal{R}\right) -\frac{\lambda ^{2}}{4}%
(D_{\mu }\vec{\phi}\times D_{\nu }\vec{\phi})^{2}-\mathcal{V}(\phi _{n})%
\text{,}  \label{L02}
\end{equation}%
where we have defined
\begin{equation}
\mathcal{R}=\sqrt{1+\frac{F_{\mu \nu }^{2}}{2\beta ^{2}g^{2}}}\text{, \ \ \
\ }\mathcal{V}(\phi _{n})=\beta ^{2}[1-U\left( \phi _{n}\right) ]\text{.}\label{vv01}
\end{equation}%
The first contribution in (\ref{L02}) is
the BI term {with $F_{\mu \nu }=\partial _{\mu }A_{\nu }-\partial
_{\nu }A_{\mu }$ is field strength tensor of the $U(1)$ gauge field $A_{\mu
} $, the electromagnetic coupling constant is given by $g$, and $\beta $ the
BI parameter. The second contribution is the Skyrme term minimally coupled
to the U(1) gauge field through the covariant derivative given by}
\begin{equation}
D_{\mu }\vec{\phi}=\partial _{\mu }\vec{\phi}+A_{\mu }\vec{n}\times \vec{\phi%
}\text{,}
\end{equation}%
where the triplet of real scalar fields $\vec{\phi}\in $ $\mathbb{S}^{2}$
represents the Skyrme field, satisfying the constraint $\vec{\phi}\cdot \vec{%
\phi}=1$. {The unitary vector $\vec{n}$ provides a preferred
direction in the internal space that afterwards becomes the vacuum value of
the Skyrme field, that is
\begin{equation}
\lim_{\left\vert \vec{x} \right\vert \rightarrow\infty} \vec{%
\phi}=\vec{n}.
\end{equation}
{Lastly, the third term in (\ref{L02}), $\mathcal{V}%
(\phi_{n})= \mathcal{V}(\vec{n}\cdot \vec{\phi})$ is the potential defined
in (\ref{vv01}) with the function $U(\phi _{n})$ satisfying the condition $%
0<U(\phi _{n})< 1$.}

We assume all the coupling constants are nonnegative quantities. Moreover,
the Skyrme field is dimensionless, the gauge field has mass dimension equal
to $1$, {both the BI parameter $\beta$ and the electromagnetic
coupling constant $g$ have mass dimension $1$, and the Skyrme coupling constant $%
\lambda $ has mass dimension $-1$.}

We now proceed to present the Euler-Lagrange equations resulting from
Lagrangian density (\ref{L02}). The equation of motion of the gauge field
reads
\begin{equation}
\partial _{\sigma }\left( \frac{1}{\mathcal{R}}F^{\sigma \mu }\right) =g^{2}{%
\ j}^{\mu }\text{,}  \label{gauge2}
\end{equation}%
{where ${j}^{\mu }=\vec{n}\cdot \vec{J}^{\mu }$ is the conserved current
density with}
\begin{equation}
\vec{J}^{\mu }=\lambda ^{2}[\vec{\phi}\cdot (D^{\mu }\vec{\phi}\times D^{\nu
}\vec{\phi})](D_{\nu }\vec{\phi})\text{.}  \label{curr}
\end{equation}%
Already for the Skyrme field we obtain
\begin{equation}
D_{\mu }\vec{J}^{\mu }=-\left\{ \lambda ^{2}\partial _{\mu }\left[ (\vec{n}%
\cdot \partial ^{\mu }\vec{\phi})\Psi ^{2}\right] +\frac{\partial \mathcal{V}%
}{\partial \phi _{n}}\right\} (\vec{n}\times \vec{\phi})\text{.}
\label{skky}
\end{equation}

Our effort will be focused to the study of stationary solutions. Then, from
Eq. (\ref{gauge2}), the Gauss law reads as
\begin{equation}
\partial _{i}\left( \frac{1}{\mathcal{R}}{\partial _{i}A_{0}}\right) =g^{2}{%
\ \lambda ^{2}A_{0}(\vec{n}\cdot \partial _{i}\vec{\phi})^{2}}\text{.}
\label{gL}
\end{equation}
{We observe that the gauge condition $A_{0}=0$ identically
satisfies the Gauss law, implying the resulting configurations only carry on
magnetic flux.}

Also, from Eq. (\ref{gauge2}), the stationary Amp\`{e}re law gives
\begin{equation}
\partial _{i}\left( \frac{B}{\mathcal{R}}\right) +\lambda ^{2}g^{2}(\vec{n}%
\cdot \partial _{i}\vec{\phi})Q=0\text{,}  \label{aL}
\end{equation}%
where $B=F_{12}=\epsilon_{ij}\partial_i A_j$ is the magnetic
field, {$\mathcal{R}$ assumes the form}
\begin{equation}
\mathcal{R}=\sqrt{1+\frac{B^{2}}{\beta ^{2}g^{2}}}\text{,}  \label{Rr}
\end{equation}%
and $Q\equiv \vec{\phi}\cdot (D_{1}\vec{\phi}\times D_{2}\vec{\phi})$, which
can still be expressed as
\begin{equation}
Q=\vec{\phi}\cdot (\partial _{1}\vec{\phi}\times \partial _{2}\vec{\phi}%
)+\epsilon _{ij}A_{i}(\vec{n}\cdot \partial _{j}\vec{\phi})\text{.}
\label{Q1}
\end{equation}%
The term $\vec{\phi}\cdot (\partial _{1}\vec{\phi}\times \partial _{2}\vec{%
\phi})$ is related to the topological charge or topological degree (also
named winding number) of the {Skyrme field},
\begin{equation}
\deg [\vec{\phi}]=-\frac{1}{4\pi }\!\int d^{2}\mathbf{x}\;\vec{\phi}\cdot
(\partial _{1}\vec{\phi}\times \partial _{2}\vec{\phi})=k\text{,}  \label{wn}
\end{equation}%
being $k\in \mathbb{Z}\setminus 0$.

Similarly, from (\ref{skky}), the stationary equation of motion of the
Skyrme field becomes
\begin{equation}
\frac{\partial U}{\mathcal{\partial }\phi _{n}}(\vec{n}\times \vec{\phi}%
)+\lambda ^{2}\epsilon _{ij}D_{i}(QD_{j}\vec{\phi})=0\text{.}  \label{wm}
\end{equation}

{In the next section, we will show how is implemented the BPS
formalism allowing us to obtain the energy lower bound and the self-dual
equations to be satisfied by the soliton configurations saturating such
a bound.}

\section{The BPS framework \label{sB}}

The stationary energy density of the model (\ref{L02}) is
\begin{equation}
\varepsilon =\beta ^{2}\left( \mathcal{R}-U\right) +\frac{\lambda ^{2}}{2}%
Q^{2}\text{,}  \label{energy0}
\end{equation}%
where we have used $Q^{2}=\frac{1}{2}(D_{i}\vec{\phi}\times D_{j}\vec{\phi}%
)^{2}$. The requirement that the energy density be null when $\left\vert
\vec{x}\right\vert \rightarrow \infty $ establishes the boundary conditions
satisfied by the fields of the model. {Thus, from (\ref{Rr}) the
magnetic field must satisfy $\lim_{\left\vert \vec{x}\right\vert \rightarrow
\infty }B=0$ implying in $\lim_{\left\vert \vec{x}\right\vert \rightarrow
\infty} \mathcal{R}=1$, consequently,
\begin{equation}
\lim_{\left\vert \vec{x}\right\vert \rightarrow \infty }U=1\text{, }%
\lim_{\left\vert \vec{x}\right\vert \rightarrow \infty }Q=0.  \label{UQ}
\end{equation}
}

The total energy is defined by integrating the energy density (\ref{energy0}%
), so that we implement the BPS formalism by writing
\begin{eqnarray}
E &=&\int d^{2}\mathbf{x}\left[ \frac{\left( B\pm \lambda ^{2}g^{2}\mathcal{R} W\right) ^{2}}{2g^{2}\mathcal{R}}+\frac{\lambda ^{2}}{2}\left( Q\mp Z\right) ^{2}\right.  \notag \\[0.2cm]
&&\hspace{1.2cm}\mp \lambda ^{2}BW\pm \lambda ^{2}QZ-\beta ^{2}U+\beta ^{2}%
\mathcal{R}  \notag \\[0.2cm]
&&\hspace{1.2cm}\left. -\frac{B^{2}}{2g^{2}\mathcal{R}}-\frac{\lambda
^{4}g^{2}\mathcal{R}W^{2}}{2}-\frac{\lambda ^{2}}{2}Z^{2}\right] ,\quad \quad
\label{en1}
\end{eqnarray}%
where we have introduced two auxiliary functions, namely $W\equiv W(\phi
_{n})$ and $Z\equiv Z(\phi _{n})$ which we shall determine later. {By using the magnetic field definition and the expressions (\ref%
{Rr}) and (\ref{Q1}), for $\mathcal{R}$ and $Q$, respectively, we arrive at}
\begin{eqnarray}
E &=&\int d^{2}\mathbf{x}\left[ \frac{\left( B\pm\lambda ^{2}g^{2} \mathcal{R}W\right) ^{2}}{2g^{2}\mathcal{R}}+\frac{\lambda ^{2}}{2}\left( Q\mp
Z\right) ^{2}\right.  \notag \\[0.2cm]
&&\hspace{-0.5cm}\pm \lambda ^{2}Z\vec{\phi}\cdot (\partial _{1}\vec{\phi}%
\times \partial _{2}\vec{\phi})\pm \lambda ^{2}\epsilon _{ij}\left[
(\partial _{j}A_{i})W+A_{i}Z(\partial _{j}\phi _{n})\right]  \notag \\[0.2cm]
&&\hspace{-0.5cm}\left. -\beta ^{2}U+\frac{\beta ^{2}}{2}\left( \frac{1}{%
\mathcal{R}}+\mathcal{R}\right) -\frac{\lambda ^{4}g^{2}\mathcal{R}W^{2}}{2}-%
\frac{\lambda ^{2}}{2}Z^{2}\right] \text{.}  \label{enNN}
\end{eqnarray}%
{At this point, to continue with the implementation of the BPS
formalism, we perform two steps: (i) the term $(\partial
_{j}A_{i})W+A_{i}Z(\partial _{j}\phi _{n})$ is transformed in a total
derivative by setting}
\begin{equation}
Z=\frac{\partial W}{\partial \phi _{n}},
\end{equation}%
and (ii) we require the function $U(\phi _{n})$ to be defined as
\begin{equation}
\beta ^{2}U=\frac{\beta ^{2}}{2}\left( \frac{1}{\mathcal{R}}+\mathcal{R}%
\right) -\frac{\lambda ^{4}g^{2}\mathcal{R}W^{2}}{2}-\frac{\lambda ^{2}}{2}%
\left( {\frac{\partial W}{\partial \phi _{n}}}\right) ^{2} \!\!.  \label{Vv}
\end{equation}
This way, the total energy becomes%
\begin{eqnarray}
E &=&\int d^{2}\mathbf{x}\left[ \frac{\left( B\pm \lambda ^{2}g^{2}\mathcal{R} W\right) ^{2}}{2g^{2}\mathcal{R}}+\frac{\lambda ^{2}}{2}\left( Q\mp {
\frac{\partial W}{\partial \phi _{n}}}\right) ^{2}\right.  \notag \\[0.3cm]
&&\quad\left. \pm \lambda ^{2}\left( {\frac{\partial W}{\partial \phi _{n}}}
\right) \vec{\phi}\cdot (\partial _{1}\vec{\phi}\times \partial _{2}\vec{\phi%
})\pm \epsilon _{ij}\partial _{j}(A_{i}W)\right] \!.\quad\quad  \label{enNN0}
\end{eqnarray}

Notably, $W(\phi _{n})$ plays the role of a superpotential function, and it
must be constructed (or proposed) such that the potential $U(\phi _{n})$ becomes unit} when $\phi _{n}\rightarrow 1$ (or $\left\vert
\vec{x}\right\vert \rightarrow \infty $) in accordance with Eq. (\ref{UQ}).
Consequently, the following boundary conditions must be satisfied,
\begin{equation}
\lim_{\phi _{n}\rightarrow 1}{W}{(\phi _{n})}=0,\ \ \lim_{\phi
_{n}\rightarrow 1}\frac{\partial W}{\partial \phi _{n}}=0.  \label{BcVv}
\end{equation}%
Then, under such boundary conditions we observe the contributions of the
total derivative in the second row of Eq. (\ref{enNN0}) vanishes. Therefore,
we can express the total energy as
\begin{equation}
E=\bar{E}+E_{_{\text{BPS}}}\text{,}  \label{en5}
\end{equation}%
where $\bar{E}$ represents {the integral composed} by the quadratic terms,
\begin{equation}
\bar{E}=\int d^{2}\mathbf{x}\left[ \frac{\left( B\pm \lambda ^{2}g^{2}\mathcal{R}W\right) ^{2}}{2g^{2}\mathcal{R}}+\frac{\lambda ^{2}}{2}\left( Q\mp {\frac{\partial W}{\partial \phi _{n}}}\right) ^{2}\right] \text{,}
\end{equation}%
and $E_{_{\text{BPS}}}$ defines the energy lower bound,
\begin{equation}
E_{_{\text{BPS}}}=\pm \lambda ^{2}\int d^{2}\mathbf{x}\left( \frac{\partial W%
}{\partial \phi _{n}}\right) \vec{\phi}\cdot (\partial _{1}\vec{\phi}\times
\partial _{2}\vec{\phi})\text{.}  \label{en3}
\end{equation}

The total energy (\ref{en5}) satisfy the inequality
\begin{equation}
E\geq E_{_{\text{BPS}}}\text{,}
\end{equation}%
because $\bar{E}\geq 0$. Then, the energy lower bound will be achieved when
the fields possess configurations such that $\bar{E}=0$, i.e., the bound is
saturated when the following set of first-order differential equations are
satisfied:
\begin{equation}
B=\mp \lambda ^{2}g^{2}W\left( 1-\frac{g^{2}\lambda ^{4}W^{2}}{\beta ^{2}}%
\right) ^{-1/2},  \label{1rd0}
\end{equation}%
\begin{equation}
Q=\pm \frac{\partial W}{\partial \phi _{n}}\text{,}  \label{2rd0}
\end{equation}%
where we have used%
\begin{equation}
\mathcal{R}=\left( 1-\frac{g^{2}\lambda ^{4}W^{2}}{\beta ^{2}}\right) ^{-1/2}%
\text{,}  \label{Rf}
\end{equation}
which allows us, together with (\ref{Vv}), write the self-dual potential of the model as
\begin{equation}
\mathcal{V}(\phi _{n})=\beta ^{2}\left[ 1-\left( 1-\frac{g^{2}\lambda
	^{4}W^{2}}{\beta ^{2}}\right) ^{1/2}\right] +\frac{\lambda ^{2}}{2}\left( {%
	\frac{\partial W}{\partial \phi _{n}}}\right) ^{2}\text{.}  \label{Ubps}
\end{equation}
The relations (\ref{1rd0}) and (\ref{2rd0}) are the called self-dual or BPS
equations which ensures the energy lower bound and stability of the field
configurations. Further, we highlight that such first order equations satisfy the
Euler-Lagrange equations associated with Lagrangian density {(\ref{L02})}.

Before proceeding for the next section, we must highlight that the BPS
configurations for the corresponding standard case {(gauged BPS baby Skyrme model)} {can be recovered in the limit $\beta \rightarrow
\infty $. {In this limit,} the BPS model described by the Lagrangian density (\ref{L02}%
) becomes}
\begin{equation}
\mathcal{L}=-\frac{1}{4g^{2}}F_{\mu \nu }^{2}-\frac{\lambda ^{2}}{4}(D_{\mu }%
\vec{\phi}\times D_{\nu }\vec{\phi})^{2}-V(\phi _{n})\text{,}  \label{L02S}
\end{equation}%
{where the corresponding BPS potential is now given by}
\begin{equation}
V(\phi _{n})=\frac{g^{2}\lambda ^{4}}{2}W^{2}+\frac{\lambda ^{2}}{2}\left( {%
\frac{\partial W}{\partial \phi _{n}}}\right) ^{2}\text{,}
\end{equation}%
{being such a system investigated in Ref. \cite{adam2}.}

\section{Rotationally Symmetric BI-Skyrmions \label{bpssolutions}}

{We now consider} solitons rotationally symmetric saturating the energy lower
bound (\ref{en3}). Henceforth, without loss of generality, we set $\vec{n}%
=(0,0,1)$ such that $\phi _{n}=\phi _{3}$ and we assume the usual \textit{%
ansatz} for the Skyrme {field},
\begin{equation}
\vec{\phi}\left( r,\theta \right) =\left(
\begin{array}{c}
\sin f(r)\cos N\theta \\
\sin f(r)\sin N\theta \\
\cos f(r)%
\end{array}%
\right)\text{,}  \label{Antz}
\end{equation}%
where $r$ and $\theta $ are polar coordinates, $N=\deg [\vec{\phi}]$ is the
winding number introduced in (\ref{wn}), and $f(r)$ a regular function
satisfying the boundary conditions
\begin{equation}
f(0)=\pi \text{,}~\ \ \lim_{r\rightarrow \infty }f(r)=0\text{.}
\label{bccr1}
\end{equation}%
{We now introduce the field {redefinition} \cite{adam2},
\begin{equation}
\phi _{3}=\cos f\equiv 1-2h\text{,}
\end{equation}%
with the field $h=h(r)$ obeying
\begin{equation}
h(0)=1\text{, \ \ }\lim_{r\rightarrow \infty }h(r)=0.  \label{bccr4}
\end{equation}%
}

For the gauge field $A_{\mu }$, we consider the {ansatz}
\begin{equation}
A_{i}=-\epsilon _{ij}{x}_{j}\frac{Na(r)}{r^{2}}\text{,}
\end{equation}%
being $a(r)$ is a behaved function satisfying the boundary condition,
\begin{equation}
\displaystyle a(0)=0,\;\lim_{r\rightarrow \infty }a(r)=a_{\infty }\text{,}
\label{bccr2}
\end{equation}%
where $a_{\infty }$ is a finite constant.

The superpotential $W(h)$ must satisfy
\begin{equation}
\lim_{r\rightarrow 0}W(h)=W_{0}\text{, \ }\lim_{r\rightarrow \infty }W(h)=0%
\text{, \ }\lim_{r\rightarrow \infty }\frac{\partial W}{\partial h}=0\text{,}
\label{c03}
\end{equation}%
with the two last conditions obtained from Eq. (\ref{BcVv}). The constant $%
W_{0}=W(h(0))=W(1)$ is a positive and finite. Besides, we consider the superpotential $W(h)$ a well-behaved function for all our analyzes.

Under the ansatz, the BPS bound (\ref{en3}) becomes
\begin{equation}
E\geq E_{_{\text{BPS}}}=\pm 2\pi \lambda ^{2}NW_{0}\text{,}  \label{en7}
\end{equation}%
being a positive-definite quantity, where the sign $+(-)$ corresponds to $%
N>0\,(N<0)$. The BPS energy density associated to this configuration may be write in form
\begin{equation}
\varepsilon _{_{\text{BPS}}}=g^{2}\lambda ^{4}W^{2}\left( 1-\frac{%
g^{2}\lambda ^{4}W^{2}}{\beta ^{2}}\right) ^{-1/2}+\frac{\lambda ^{2}}{4}%
\left( \frac{\partial W}{\partial h}\right) ^{2}\text{.}
\end{equation}

Similarly, the BPS equations (\ref{1rd0}) and (\ref{2rd0})
assume, respectively, the form
\begin{equation}
\frac{N}{r}\frac{da}{dr}+\lambda ^{2}g^{2}W\left( 1-\frac{g^{2}\lambda
^{4}W^{2}}{\beta ^{2}}\right) ^{-1/2}=0\text{,}  \label{bc01}
\end{equation}%
\begin{equation}
\frac{4N}{r}\left( 1+a\right) \frac{dh}{dr}+\frac{\partial W}{\partial h}=0%
\text{,}  \label{bc02}
\end{equation}%
where has been used the magnetic field given by%
\begin{equation*}
B=\frac{N}{r}\frac{da}{dr}\text{.}
\end{equation*}%
Also, note that we have chosen, without a loss of generality, the upper
sign. Such an assumption will be considered in the remaining of the
manuscript.

In what follows, we will present the behavior of the self-dual profiles close to the boundaries by solving the BPS equations (\ref{bc01}) and (\ref{bc02}) according to the already established boundary conditions. We begin showing the behavior of the fields around the origin,  which are given by
\begin{align}
h(r)& \approx 1-\frac{\left( W_{h}\right) _{h=1}}{8N}r^{2}+\frac{\left(
W_{h}\right) _{h=0}\left( W_{hh}\right) _{h=1}}{128N^{2}}r^{4}\text{,}\label{ag01} \\%
[0.2cm]
a(r)& \approx -\frac{\lambda ^{2}g^{2}\mathcal{A}_{0}W_{0}}{2N}r^{2}+\frac{%
\lambda ^{2}g^{2}\mathcal{A}_{0}^{3}\left( W_{h}\right) _{h=1}^{2}}{32N^{2}}%
r^{4}\text{,}\label{ag02}
\end{align}%
where $W_{h}={\partial W}/{\partial h}$, $W_{hh}={\partial ^{2}W}/{\partial
h^{2}}$ {and the constant }$\mathcal{A}${$_{0}$ defined as}%
\begin{equation}
\mathcal{A}_{0}=\left( 1-\frac{g^{2}\lambda ^{4}W_{0}^{2}}{\beta ^{2}}%
\right) ^{-1/2}\text{.} \label{AA0}
\end{equation}
Furthermore, near the origin, for the magnetic field and BPS energy density we get, respectively,

\begin{equation}
B(r)\approx -\lambda ^{2}g^{2}\mathcal{A}_{0}W_{0}+\frac{\lambda ^{2}g^{2}%
\mathcal{A}_{0}^{3}\left( W_{h}\right) _{h=1}^{2}}{8N}r^{2}\text{,} \label{B0}
\end{equation}
and
\begin{equation}
\varepsilon _{_{\text{BPS}}}\approx \frac{\lambda ^{2}(W_{h})_{h=1}^{2}}{4}%
+\lambda ^{4}g^{2}\mathcal{A}_{0}W_{0}^{2} -\frac{\lambda ^{4}g^{2}\mathcal{A%
	}_{1}(W_{h})_{h=1}^{2}}{8N} r^{2} ,  \label{B1}
\end{equation}
where $\mathcal{A}_{1}$ has been defined as
\begin{equation}
\mathcal{A}_{1}=\frac{(W_{hh})_{h=1}}{2\lambda ^{2}g^{2}}+ (1+\mathcal{A}%
_{0}^{2}) \mathcal{A}_{0}W_{0}.
\end{equation}

{A brief analysis mainly from the behavior near the origin of
	the gauge field profile (\ref{ag01}), magnetic field (\ref{B0}) and BPS
	energy density (\ref{B1}) reveals the existence of a singularity associated
	with the BI parameter as shown by the constant $\mathcal{A}_{0}$ defined in
	Eq. (\ref{AA0}), i.e., the field profiles will be well-defined if and only
	if the $\beta$-parameter satisfies}
\begin{equation}
\beta>\beta _{c}=g\lambda ^{2}W_{0}\text{.}  \label{betaP}
\end{equation}%
Thus, we expect that well-behaved solitons should exist in
the range $\beta _{c}<\beta<\infty $ and that for $\beta $ sufficiently
large, the profiles become similar to the those engendered by the model (\ref{L02S}).

On the other hand, to compute the field profiles behavior for sufficiently large values of $r$, we consider a superpotential $W(h)$ behaving as
\begin{equation}
W(h)\approx W_{R}^{(\sigma)}h^{\sigma }\;\;\text{with}\;\;\sigma >{1},\label{Whv}
\end{equation}%
where $W_{R}^{(\sigma)}>0$. The asymptotic form in Eq. (\ref{Whv}) engenders potentials  (\ref{Ubps}) behaving as $\mathcal{V} \sim h^{2\sigma-2}$, allowing us to compare our solutions with those of the cases studied in  Ref. \cite{adam2}.

Our analysis is performed by considering the boundary conditions
\begin{equation}
h (R) =0\text{, \ \ \ }a(R) =a_{R}\text{, \ \ \ \ }W(h(R))=0\text{,}\label{bcR}
\end{equation}%
with $R>0$ and $a_{R}$ being a real constant. {A finite value of $R$ defines the maximum size of the topological defect characterizing a soliton named compacton, i.e., it achieves the vacuum value in a finite radius $R$ (the compacton's radius) and remains in the vacuum for all $r> R$. On the other hand,  when $R\rightarrow \infty $ the model engenders extended or noncompact configurations that may be localized or delocalized. Hence, we have different solitons configurations satisfying the boundary conditions (\ref{bcR}).  Of course, these solutions depending on the $\sigma$ values; this way, the asymptotic analysis leads us to three types of solitons: (i) for $1<\sigma<2$ we have compactons; (ii) for $\sigma =2$ the soliton tail decays following a Gaussian-law;  (iii) for $\sigma >2$, the soliton tails have a power-law decay. Besides, all these soliton profiles behave near to the origin according as already presented in (\ref{ag01}) and (\ref{ag02}).}

{We now calculate the magnetic flux of the BPS solitons. It is given by
\begin{equation}
\Phi =2\pi \int_0^R B\, rdr=2\pi Na_{R} \label{FlxB}
\end{equation}
for compactons, whereas for the noncompact solitons, the magnetic flux is
\begin{equation}
\Phi =2\pi \int_0^\infty B\, rdr=2\pi Na_{\infty}. \label{FlxB1}
\end{equation}
We observe the magnetic flux is nonquantized (in the topological sense) since $a_R$ or $a_{\infty}$ are real numbers belonging to the interval $\left<-1,0\right]$. However, for sufficiently large values of $g$, the vacuum values $a_R$ or $a_{\infty}$ tends to -1; consequently, in such a limit, the magnetic flux becomes quantized in units of $2\pi$.}

In the following sections, we will show the field profiles behavior in the limit $r\rightarrow\infty$ and also perform the numerical solution of the BPS equations (\ref{bc01}) and (\ref{bc02}). Thus, we will choose specific superpotentials for such aims, allowing us to study the different types of solutions mentioned previously.

\subsection{Born-Infeld compactons}

A superpotential like (\ref{Whv}) behaving  for $r\rightarrow R$, i.e.,
\begin{equation}
W(h)\approx W_{R}h^{\sigma } \quad \text{with}\quad  1<\sigma <2,  \label{Whvc}
\end{equation}
engenders Born-Infeld compactons whose profiles possess the following behavior:

\begin{equation}
h(r)\approx \mathcal{C}_{R}\left( R-r\right) ^{\frac{1}{2-\sigma }}+%
\frac{\mathcal{C}_{R}}{2\left( 2-\sigma \right) R}\left( R-r\right)
^{\frac{3-\sigma }{2-\sigma }}\text{,}
\end{equation}%
\begin{align}
a(r)& \approx a_{R}+\frac{2{g}^{2}{\lambda }^{2}(1+a_{R})\left( \mathcal{C}_{R}\right) ^{2}}{%
	\sigma }\left( R-r\right) ^{\frac{2}{2-\sigma }}+\cdots  \notag \\[0.2cm]
& +\,{\frac{{g}^{4}{%
			\lambda }^{6}{W_{{R}}}^{2}\left( 1+a_{R}\right) }{{\beta }^{2}\sigma \left(
		1+\sigma \right) }}\left( \mathcal{C}_{R}\right) {^{2+2\sigma }}\left(
R-r\right) ^{\frac{2+2\sigma }{2-\sigma }}\text{,}
\end{align}%
{where we have considered the lowest-order in $R-r$ and the
first-contribution of the BI parameter. We also have} defined the quantity $\mathcal{C}_{R}$ as
\begin{equation}
\mathcal{C}_{R}=\left[ \frac{R\sigma \left( 2-\sigma
\right) W_{R}}{4N\left( a_{R}+1\right) }\right] ^{1/\left( 2-\sigma \right) }%
\text{,}  \label{expsq}
\end{equation}%
{being $a_{R}=a(R)$ the} vacuum value of the gauge field profile.

The magnetic field and BPS energy density have their first relevant terms given by
\begin{eqnarray}
B &=&-W_{{R}}{g}^{2}{\lambda }^{2}\left( \mathcal{C}_{R}\right) ^{\sigma
}\left( R-r\right) ^{\frac{\sigma }{2-\sigma }}+\cdots   \notag \\
&&-{\frac{{g}^{4}{\lambda }^{6}{W_{{R}}}^{3}}{2{\beta }^{2}}}\left( \mathcal{%
	C}_{R}\right) {^{3\sigma }}\left( R-r\right) ^{\frac{3\sigma }{2-\sigma }}\text{,}
\end{eqnarray}
and
\begin{eqnarray}
\varepsilon _{_{\text{BPS}}} &=&\frac{W_{{R}}^{2}{\lambda }^{2}}{4}\sigma
^{2}\left( \mathcal{C}_{R}\right) ^{2\sigma -2}\left( R-r\right) ^{\frac{2\sigma -2}{2-\sigma }}+\cdots   \notag \\
&&+\,{\frac{{g}^{4}{\lambda }^{8}{W_{{R}}}^{4}\left( 2-\sigma \right)}{2{\beta }^{2}\left( 1+\sigma \right) }}\left( \mathcal{C}_{R}\right) {^{4\sigma }} \left( R-r\right) ^{\frac{4\sigma }{2-\sigma }}\text{,}
\end{eqnarray}
respectively.

\begin{figure}[t]
\includegraphics[width=4.15cm]{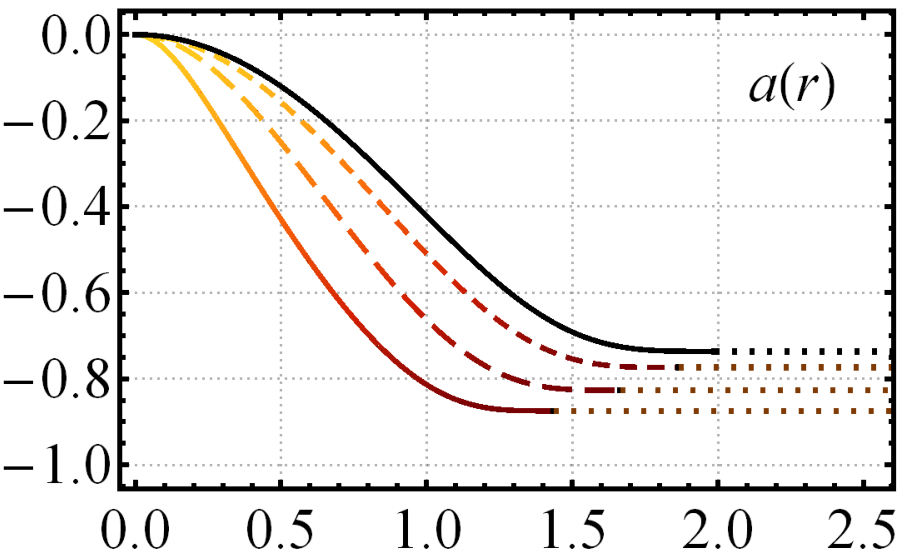}\hspace{0.1cm} %
\includegraphics[width=4.15cm]{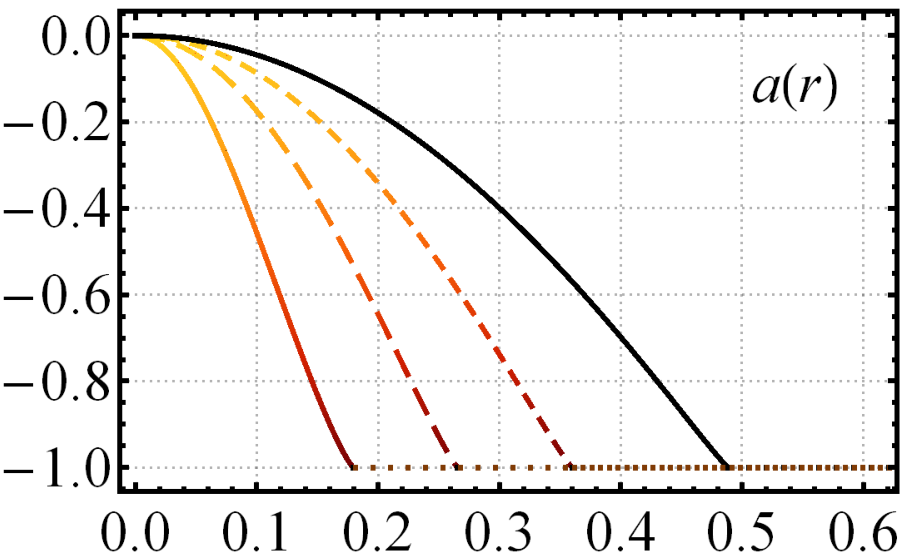}\\ \vspace{0.2cm}
\includegraphics[width=4.15cm]{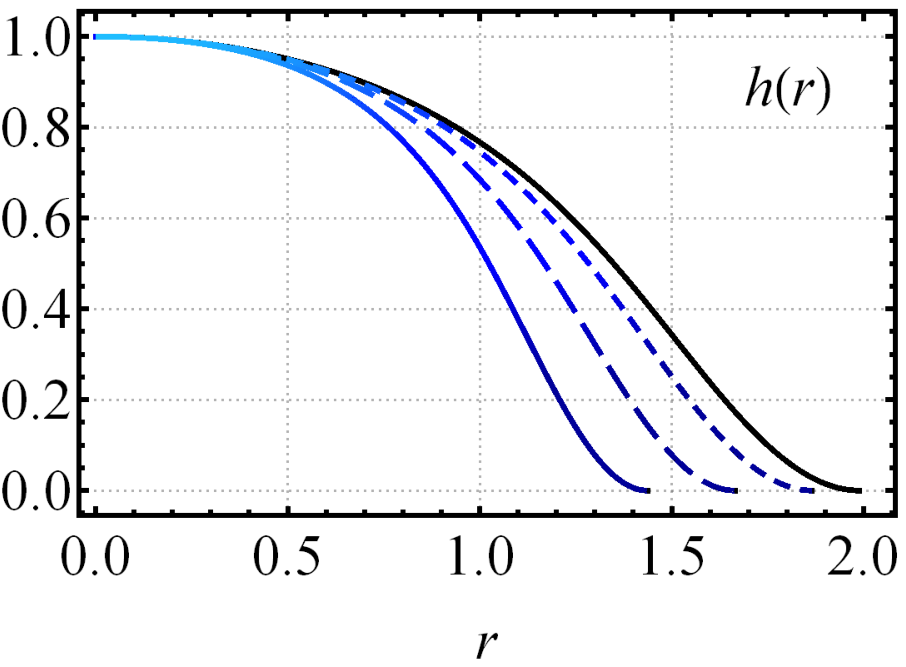}\hspace{0.1cm} %
\includegraphics[width=4.15cm]{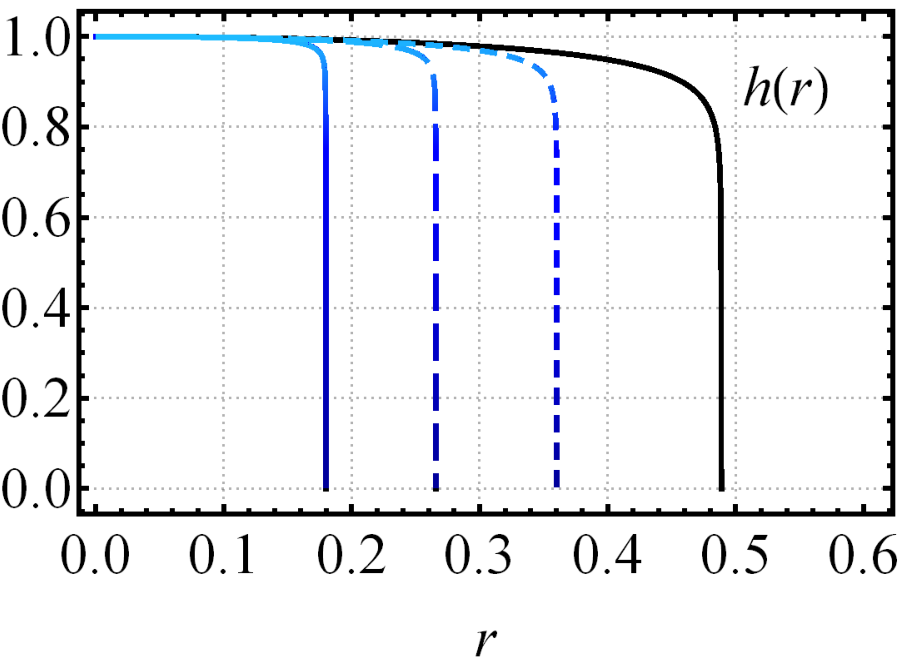}
\caption{The  compacton profiles (color lines) generated by the superpotential (\ref{Whv2}) with $g=1$ (left), $g=3$ (right),  and different values of $\beta$. Besides, we depict the corresponding ones fathered  by the model (\ref{L02S}) (solid black lines).  For the Born-Infeld Skyrme model, we depict $\beta=\beta_{c}+0.01$ (solid lines), $\beta= \beta_{c} + 0.1$ (long-dashed lines), and $\beta=\beta_{c}+0.5$ (dashed lines). {The pointed lines in the profiles for the gauge field $a(r)$ stands for $r>R$ (outside of the compacton)}.}\label{Fig01}
\end{figure}

\begin{figure}[t]
	\includegraphics[width=4.15cm]{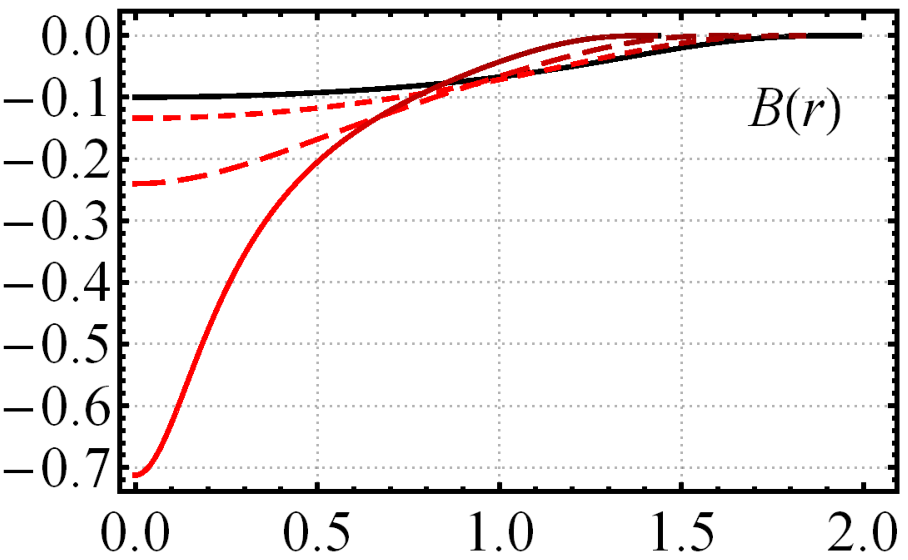}\hspace{0.1cm} %
	\includegraphics[width=4.15cm]{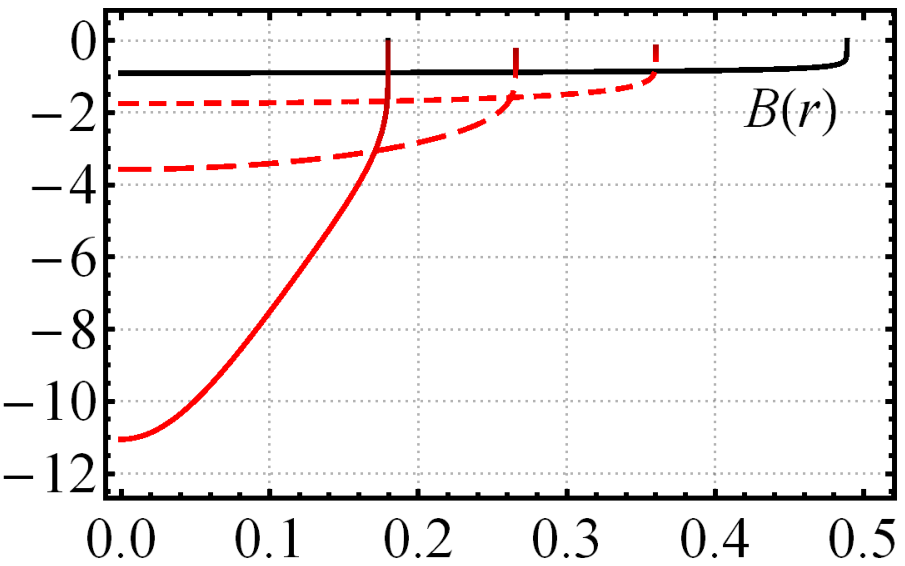}\\ \vspace{0.2cm}
	\includegraphics[width=4.15cm]{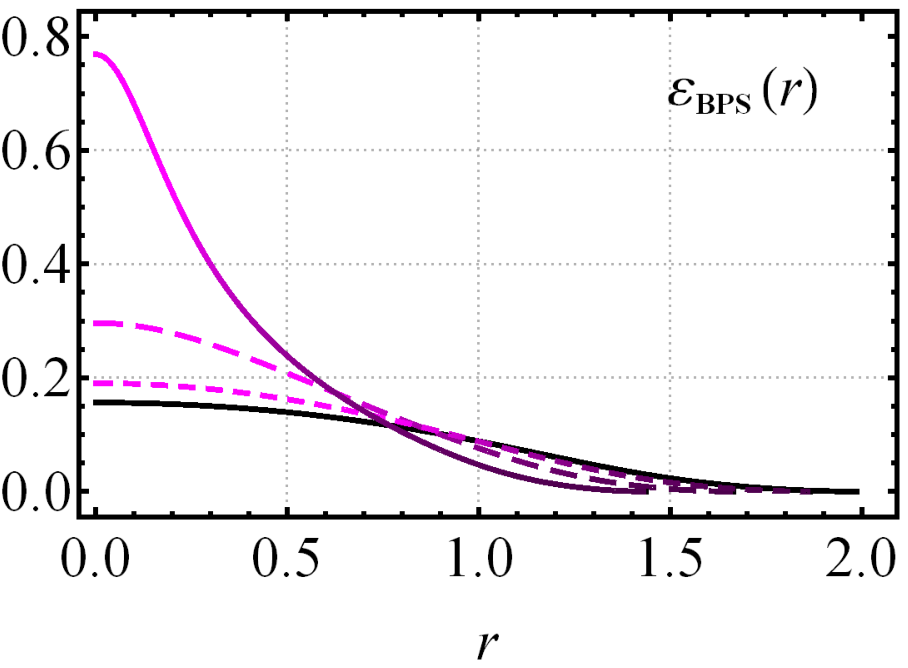}\hspace{0.1cm} %
	\includegraphics[width=4.15cm]{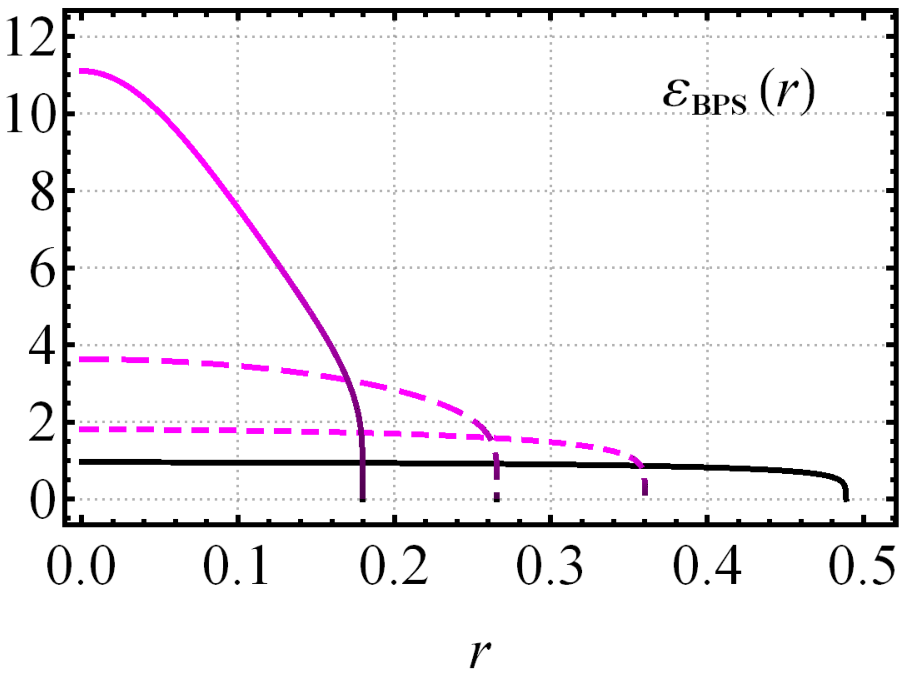}
	\caption{The profiles for magnetic field and BPS energy density
		rescaled by the factor $10^{-1}$. Conventions as in Fig. \ref{Fig01}.}
	\label{Fig02}
\end{figure}

For the numerical solutions, we select a superpotential behaving as (\ref{Whv}) setting  $\sigma=3/2$. This way, we assume
\begin{equation}
W(h)=W_0 h^{3/2} \text{,}  \label{Whv2}
\end{equation}%
where we choose $W_{0}=1/\lambda^2$. This superpotential for $r\rightarrow R$ engenders a potential behaving as $\mathcal{V}\sim h$, which is analogous to the so-called ``old baby Skyrme potential" \cite{adam2}.  For simplicity, we have fixed $N=1$, $\lambda=1$ henceforth and perform our numerical analysis in the following way:  given a coupling constant $g$, we run distinct values for the BI parameter $\beta$. The compacton solutions are depicted in Figs. \ref{Fig01} and \ref{Fig02}.

Once the parameter $\beta $ controls the BI term, which
behaves as Maxwell's term for sufficiently large values of $\beta $, we expect the compacton profiles (color lines) to be similar to those of the gauged BPS baby Skyrme model (black lines), as shown in
Figs. \ref{Fig01} and \ref{Fig02}. On the other hand, we observe the
compacton radius shrinks when $\beta$ diminishes continuously but remains
greater than the critical value $\beta_{c}$. Furthermore, we note the compacton radius diminishes when $g$ values increase (compare left and right pictures in Fig. \ref{Fig01}), a general feature already observed in the solutions of its standard counterpart. Also, Fig. \ref{Fig02} shows that when the $\beta$ values are closer to $\beta_{c}$, the profiles of the magnetic field and BPS energy density suffer relevant changes, and the respective values at the origin overgrow.

We point out that, for our analysis, we have conveniently considered the values ($g\geq 1$). It is because when $g$ is small, the nonlinear effects of the BI term becomes insignificant, which is clarified by looking in Eq. (\ref{bc01}), where the term in brackets depending on $\beta$ becomes higher-order than $g^2$ due to the constraint (\ref{betaP}). This way, for sufficiently small values of $g$, the standard case becomes dominant such as verified numerically. Similar considerations also arise for the case of the extended solitons, which we will analyze in the following sections.

\subsection{Localized Born-Infeld Skyrmions}

\label{sigma2}

{As already commented, when the vacuum value happens at $%
	R\rightarrow \infty $ is engendered noncompact configurations. This way, for
	superpotentials behaving like (\ref{Whv}) with $\sigma =2$, i.e.,}
\begin{equation}
W(h)\approx W_{\infty }^{(2)}h^{2}\text{,}  \label{superh2}
\end{equation}%
the resulting extended configurations represent localized
solitons, whose field profiles behave as
\begin{equation}
h(r)\approx \mathcal{C}_{\infty }^{(2)}e^{-\Lambda r^{2}}\text{,}
\end{equation}%
\begin{align}
a(r)& \approx a_{\infty }+\lambda ^{2}g^{2}\left( 1+a_{\infty }\right)
\left( \mathcal{C}_{\infty }^{(2)}\right) ^{2}e^{-2\Lambda r^{2}}  \notag \\%
[0.2cm]
& +\frac{g^{4}\lambda ^{6}\left( 1+a_{\infty }\right) \left( W_{\infty
}^{(2)}\right) ^{2}}{6\beta ^{2}}\left( \mathcal{C}_{\infty }^{(2)}\right)
^{6}e^{-6\Lambda r^{2}}\text{,}
\end{align}%
{where we have considered the first-contribution of the BI parameter. Besides $\mathcal{C}_{\infty }^{(2)}>0$, the parameter $a_{\infty } =a(\infty)$ is the vacuum value of the gauge field profile, and $\Lambda>0 $ is given by}
\begin{equation}
\Lambda =\frac{W_{\infty }^{(2)}}{4N\left( 1+a_{\infty }\right) }\text{.}
\end{equation}%
Further, we can exhibit the corresponding behaviors for both the magnetic field and BPS energy density, which read as
\begin{eqnarray}
B(r) &\approx &-\lambda ^{2}g^{2}W_{\infty }^{(2)}\left( \mathcal{C}_{\infty
}^{(2)}\right) ^{2}e^{-2\Lambda r^{2}}+...  \notag \\[0.2cm]
&&-\frac{g^{4}\lambda ^{6}\left( W_{\infty }^{(2)}\right) ^{3}\left(
	\mathcal{C}_{\infty }^{(2)}\right) ^{6}}{2\beta ^{2}}e^{-6\Lambda r^{2}}%
\text{,}
\end{eqnarray}%
and
\begin{eqnarray}
\varepsilon _{_{\text{BPS}}} &\approx &\lambda ^{2}\left( W_{\infty
}^{(2)}\right) ^{2}\left( \mathcal{C}_{\infty }^{(2)}\right)
^{2}e^{-2\Lambda r^{2}}+...  \notag \\[0.2cm]
&&+\frac{g^{4}\lambda ^{8}\left( W_{\infty }^{(2)}\right) ^{4}\left(
	\mathcal{C}_{\infty }^{(2)}\right) ^{8}}{2\beta ^{2}}e^{-8\Lambda r^{2}},
\end{eqnarray}%
respectively.

\begin{figure}[t]
	\includegraphics[width=4.15cm]{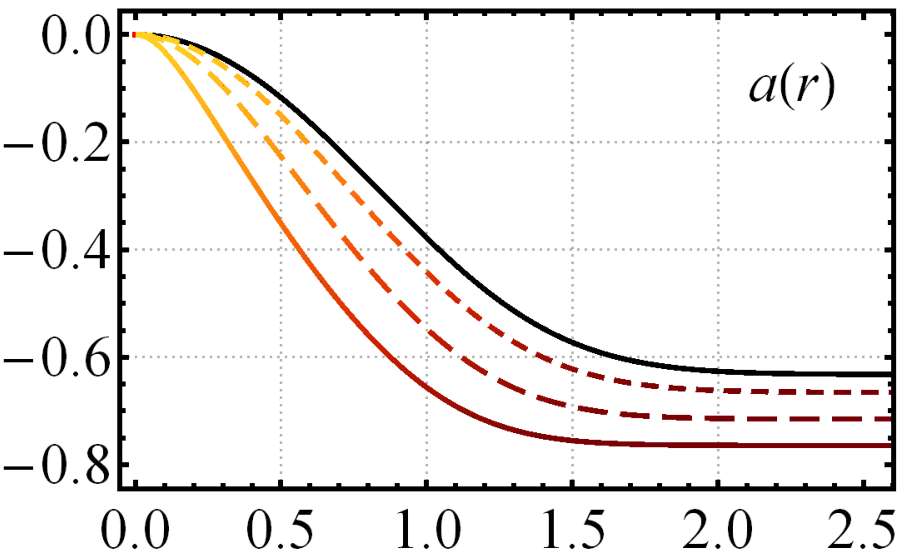}\hspace{0.1cm} %
	\includegraphics[width=4.15cm]{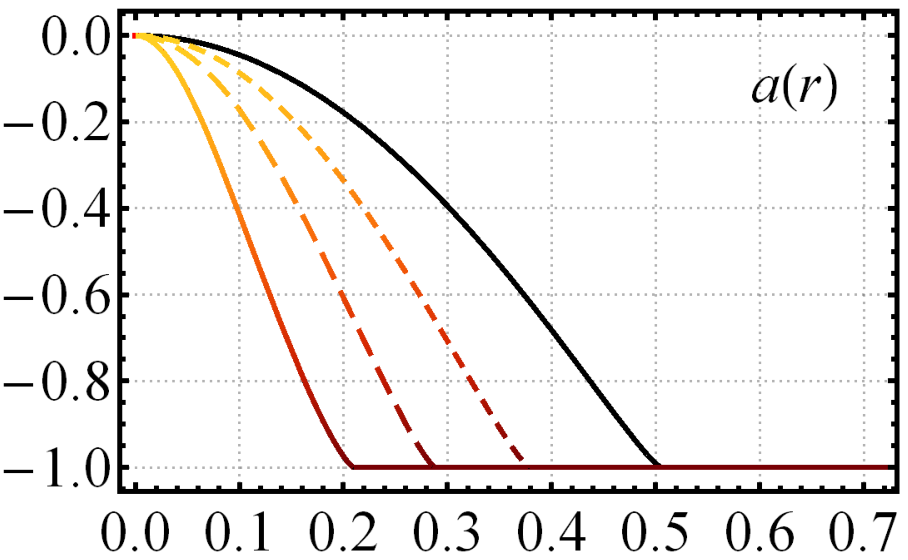}\\ \vspace{0.2cm}
	\includegraphics[width=4.15cm]{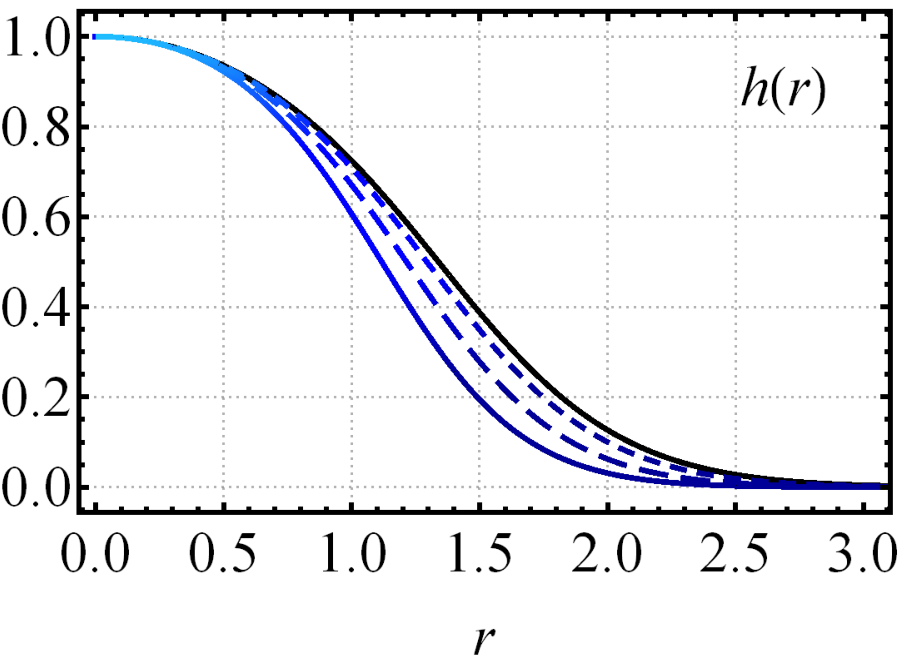}\hspace{0.1cm} %
	\includegraphics[width=4.15cm]{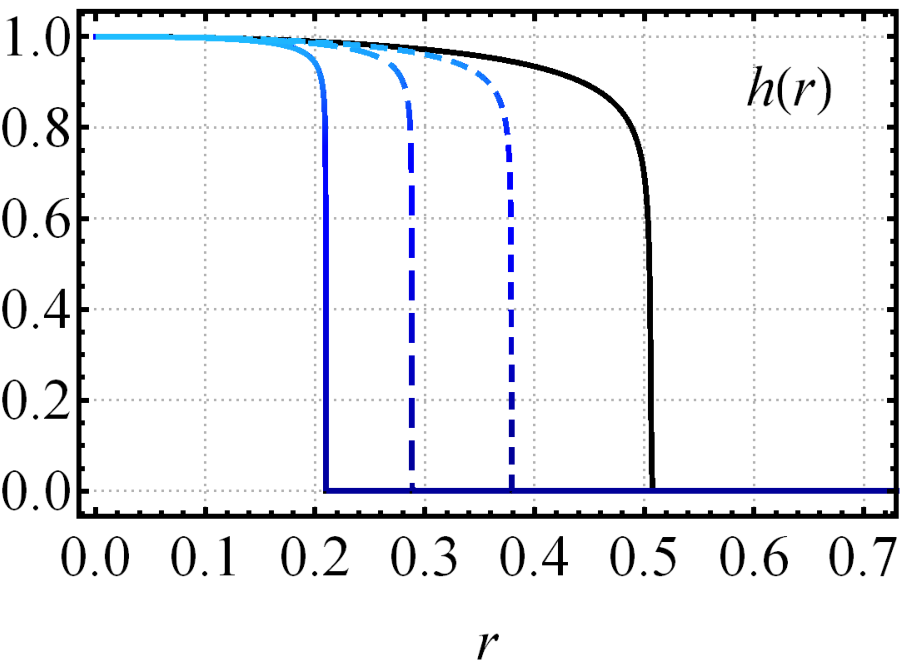}
	\caption{The profiles for the solitons by assuming the superpotential (%
		\ref{superh2a}). Conventions as in Fig. \ref{Fig01}.}
	\label{Fig03}
\end{figure}

\begin{figure}[t]
	\includegraphics[width=4.15cm]{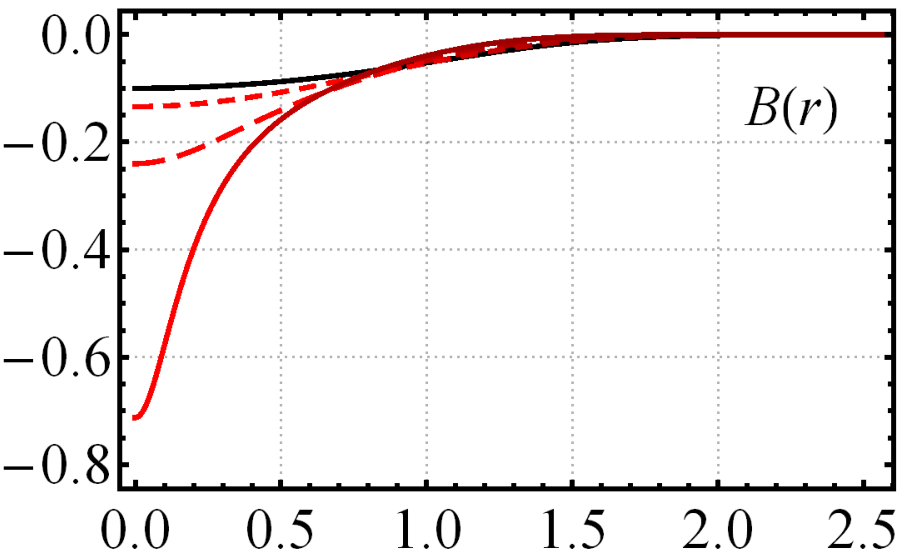}\hspace{0.1cm} %
	\includegraphics[width=4.15cm]{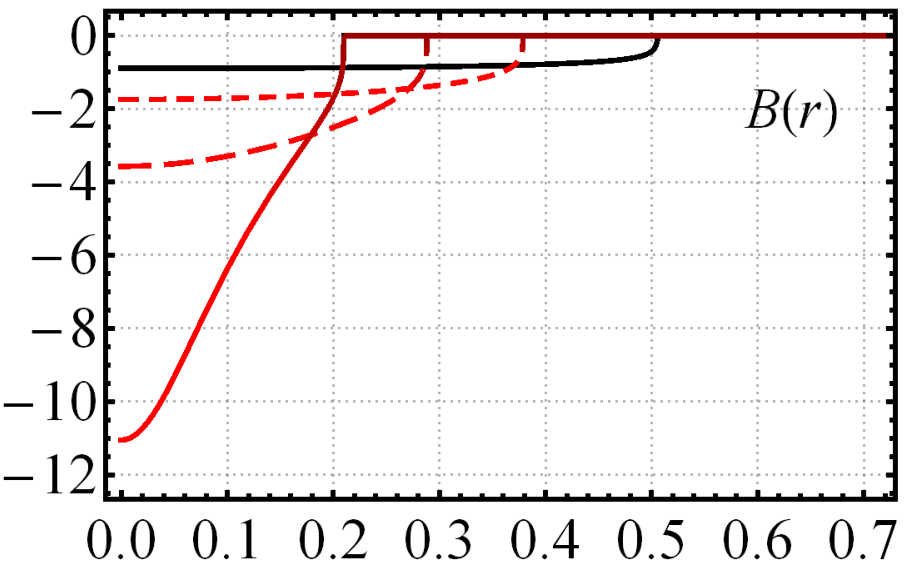}\\ \vspace{0.2cm}
	\includegraphics[width=4.15cm]{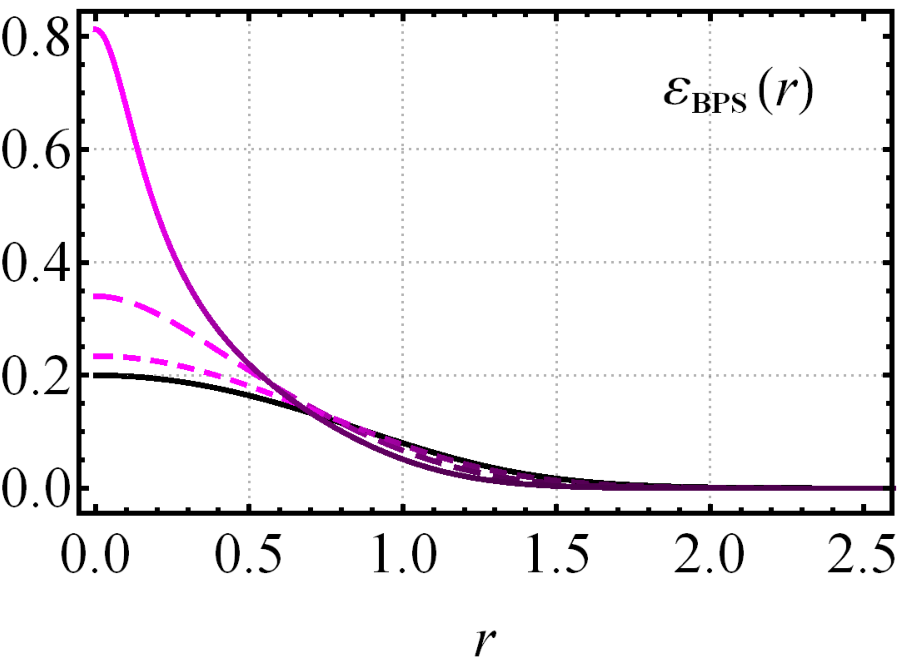}\hspace{0.1cm} %
	\includegraphics[width=4.15cm]{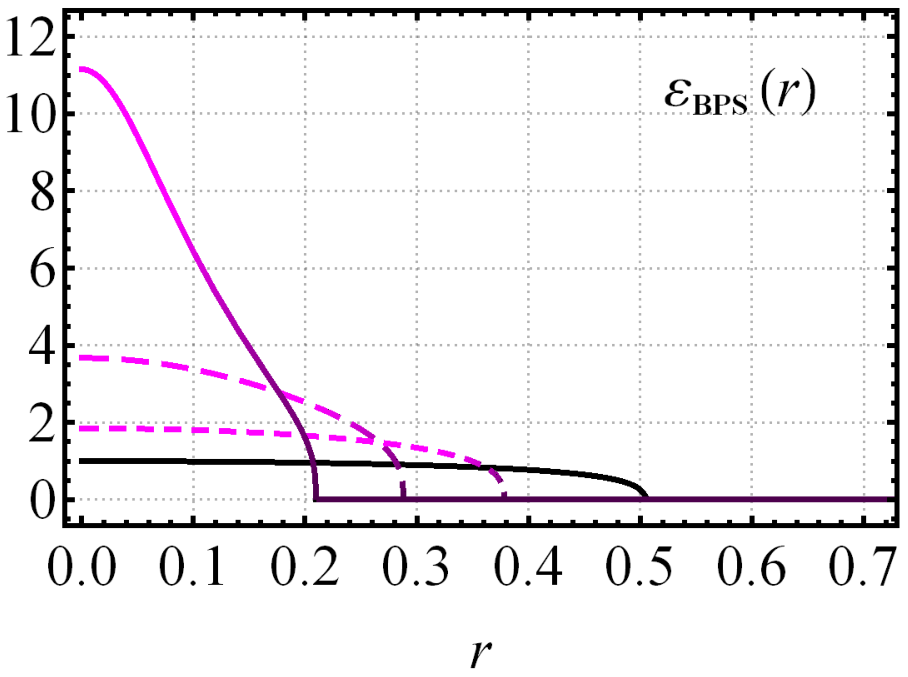}
	\caption{The profiles for the solitons by assuming the superpotential (%
		\ref{superh2a}). Conventions as in Fig. \ref{Fig02}.}
	\label{Fig04}
\end{figure}

For our numerical analysis, we have set the superpotential as
\begin{equation}
W(h)= W_0 h^{2} ,  \label{superh2a}
\end{equation}
where we  use $W_{0}=1/\lambda^2$. Here, the superpotential for $r\rightarrow\infty$ provides a potential behaving as $\mathcal{V}\sim h^2$, which also has an analogous version investigated in Ref. \cite{adam2}. As in the previous case, we adopt $N=1$ and $\lambda=1$, and we run distinct values for the BI parameter $\beta$ for a fixed value of the coupling constant $g$. The numerical solutions depicted in Figs. \ref{Fig03} and \ref{Fig04} show the field profiles, the magnetic field, and BPS energy density. Observing the profiles, we perceive they present general aspects already described previously discussed for the compacton case. However, the solitons are extended and localized whose tails, for sufficiently large values of $r$, decay following a Gaussian law.

\subsection{Delocalized Born-Infeld Skyrmions}

{We now consider a superpotential whose behavior for $%
	r\rightarrow \infty $ is given by Eq. (\ref{Whv}) with $\sigma >2$,}
\begin{equation}
W(h)\approx W_{\infty }^{(\sigma )}h^{\sigma }\text{,}  \label{sigma2m}
\end{equation}%
where $W_{\infty }^{(\sigma )}$ is a positive constant. Then, in this limit, the field
profiles have the asymptotic behavior
\begin{equation}
h(r) \approx  \left( \frac{\mathcal{C}_{\infty }^{(\sigma )}}{r^{2}}\right)
^{\frac{1}{\sigma -2}}\text{,}  \label{hsigS}
\end{equation}%
\begin{eqnarray}
a(r) &\approx &a_{\infty }+\allowbreak \frac{2{g}^{2}{\lambda }^{2}\left(
	1+a_{\infty }\right) }{\sigma }\left( \frac{\mathcal{C}_{\infty }^{(\sigma )}%
}{r^{2}}\right) ^{\frac{2}{\sigma -2}}+...  \notag \\[0.2cm]
&&+\frac{{g}^{4}{\lambda }^{6}\left( 1+a_{\infty }\right) \left( W_{\infty
	}^{(\sigma )}\right) ^{2}}{\sigma \left( \sigma +1\right) \beta ^{2}}\left(
\frac{\mathcal{C}_{\infty }^{(\sigma )}}{r^{2}}\right) ^{\frac{2\sigma +2}{%
		\sigma -2}}\text{,}
\end{eqnarray}%
{where we have considered the lowest-order in $r^{-1}$ and the
	first-contribution of the BI parameter. The constant $\mathcal{C}_{\infty
	}^{(\sigma )}$ is given by}
\begin{equation}
\mathcal{C}_{\infty }^{(\sigma )}=\frac{8N\left( a_{\infty }+1\right) }{%
	\sigma \left( \sigma -2\right) W_{\infty }^{(\sigma )}},
\end{equation}
{remembering that the parameter $a_{\infty } =a(\infty)$ is the vacuum value of the gauge field profile. Furthermore,} the magnetic field and BPS energy density behave as

\begin{eqnarray}
B(r) &\approx &-\lambda ^{2}g^{2}W_{\infty }^{(\sigma )}\left( \frac{%
	\mathcal{C}_{\infty }^{(\sigma )}}{r^{2}}\right) ^{\frac{\sigma }{\sigma -2}%
}+...  \notag \\
&&-\frac{g^{4}\lambda ^{6}\left( W_{\infty }^{(\sigma )}\right) ^{3}}{2\beta
	^{2}}\left( \frac{C_{\infty }^{(\sigma )}}{r^{2}}\right) ^{\frac{3\sigma }{%
		\sigma -2}}\text{,}
\end{eqnarray}%
and%
\begin{eqnarray}
\varepsilon _{_{\text{BPS}}}& \approx &\frac{1}{4}\sigma ^{2}\lambda
^{2}\left( W_{\infty }^{(\sigma )}\right) ^{2}\left( \frac{\mathcal{C}%
	_{\infty }^{(\sigma )}}{r^{2}}\right) ^{\frac{2\sigma -2}{\sigma -2}}+...
\notag \\[0.2cm]
&& +\frac{g^{4}\lambda ^{8}\left( W_{\infty }^{(2)}\right) ^{4}}{2\beta ^{2}}%
\left( \frac{\mathcal{C}_{\infty }^{(\sigma )}}{r^{2}}\right) ^{\frac{%
		4\sigma }{\sigma -2}}\text{,}
\end{eqnarray}%
respectively.

\begin{figure}[t]
\includegraphics[width=4.1cm]{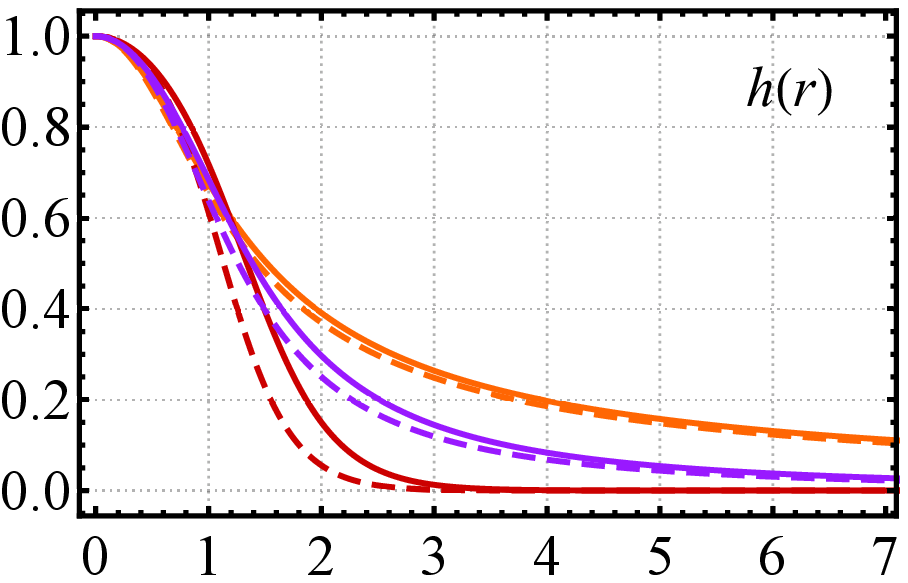} %
\includegraphics[width=4.25cm]{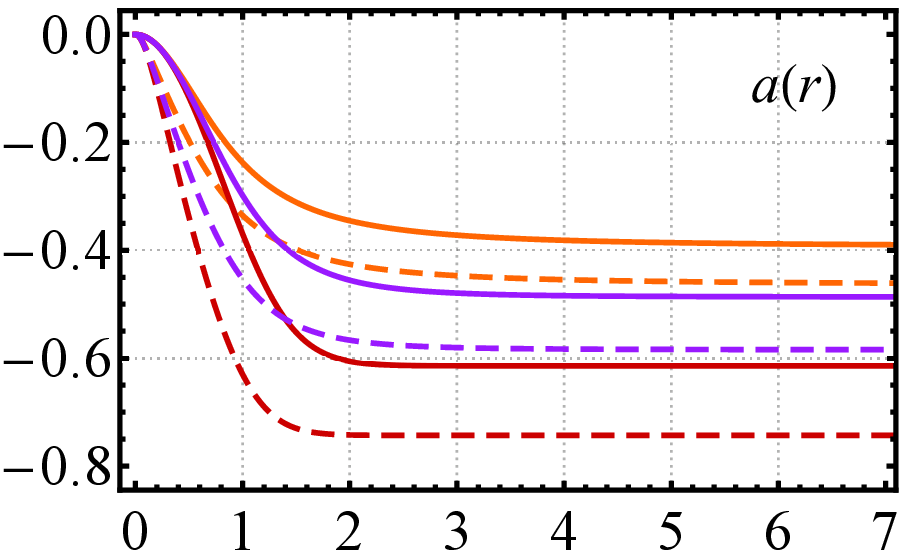}\\ \vspace{0.2cm}
\includegraphics[width=4.25cm]{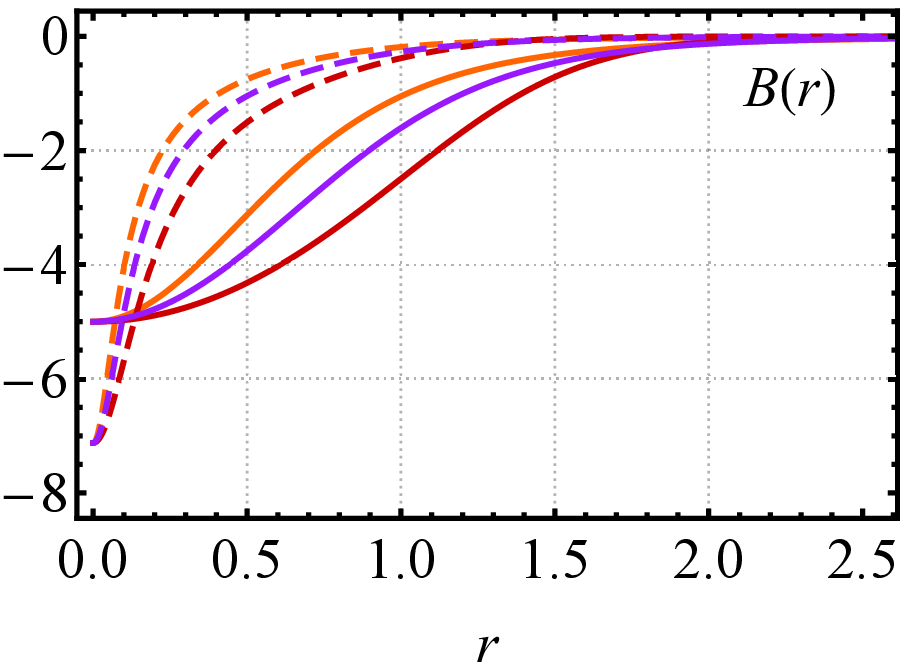}
\includegraphics[width=4.15cm]{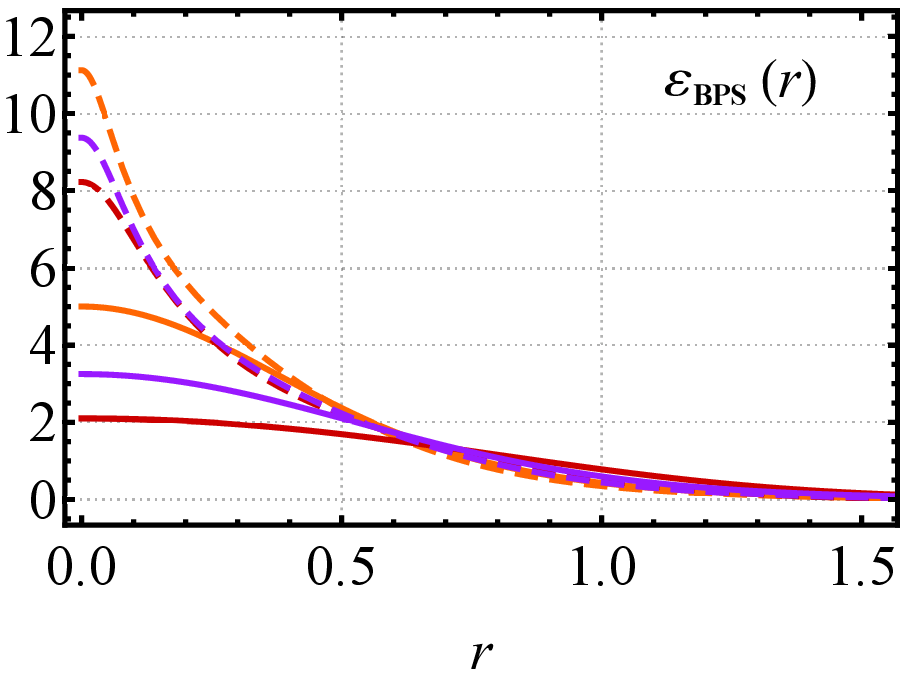}
\caption{The profiles for the solitons (dashed lines) generated by the superpotential (\ref{superh2c}) with $g=1$, $\beta=1.01$, and distinct values of $\sigma$ which are exhibited together with the ones of the counterpart model (\ref{L02S}) (solid lines). We represent the profiles with $\sigma=2.1$ (red lines), $\sigma=3.0$ (purple lines), and $\sigma=4.0$ (orange lines). The $B(r)$ profiles for the standard case has been rescaled by the factor 5.} \label{Fig05}
\end{figure}

{To perform the numerical analysis, we have selected the following superpotential
\begin{equation}
W(h)= W_0 h^{\sigma},\quad \sigma>2,  \label{superh2c}
\end{equation}
by setting $W_{0}=1/\lambda^2$.
Next, by considering $N=1$, $\lambda=1$, and fixing both the coupling constants $g$ and $\beta$, we run distinct values for the parameter $\sigma$. We have adopted a different approach than the two previous cases because we want to analyze the soliton's features for $\sigma>2$. With such a perspective, the numerical solutions depicted in Fig. \ref{Fig05} show the field profiles, the magnetic field, and the BPS energy density}.

We observe that behavior of the Skyrme field profiles $h(r)$ follow a decay more slowly to its vacuum value whenever $\sigma$ {increases by following the power-law} presented in Eq. (\ref{hsigS}).  {We also see that the Skyrme field profiles and the corresponding ones to the gauged BPS  baby  Skyrme model become} closer and closer as $\sigma$ grows. {The same effect also is} observed in the gauge field profiles $a(r)$. On the other hand, the magnetic field and BPS energy density exhibit distinct values at the origin {and different formats of the standard counterpart.}  Interestingly, {for every model,  unlike the BPS-energy density, }the magnetic field has the same values at the origin {for all values of $\sigma$. It happens because of our choice {$\lambda=1$ (hence $W_0=1$)} in all cases verifying the result shown in Eq. (\ref{B0}).}

\begin{figure}[t]
	\includegraphics[width=4.1cm]{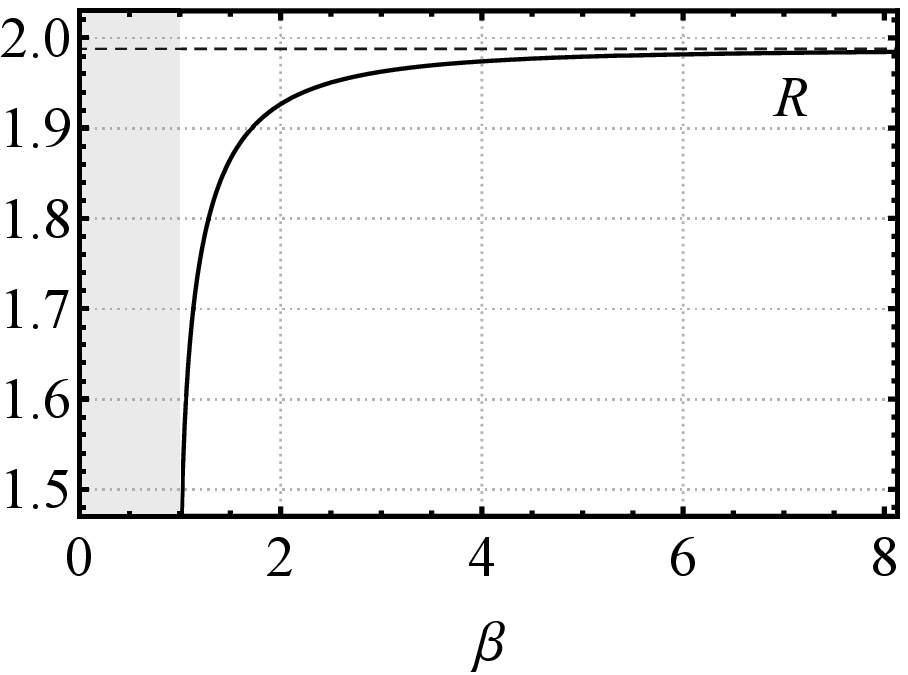}\hspace{0.1cm} %
	\includegraphics[width=4.18cm]{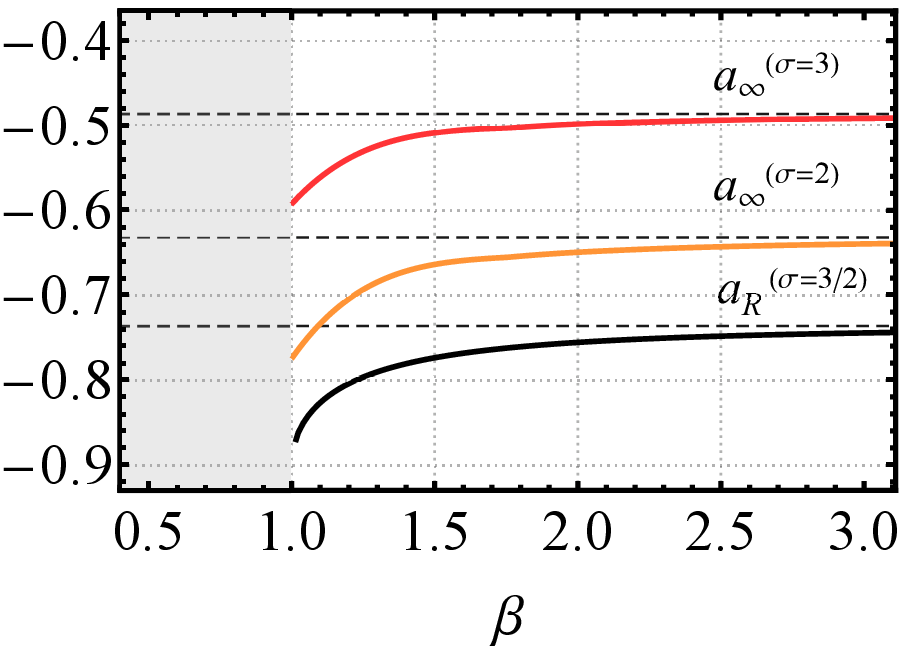}
	\caption{On the left, we plot the compacton radius $R$ vs. $\beta$. On the right-side, we have the gauge field vacuum value vs. $\beta$ for the three distinct superpotentials here analyzed. We have set $N=1$, $g=1$, and $\lambda=1$. The dashed lines depict the standard counterparts.} \label{Fig06}
\end{figure}

\begin{figure}[b]
	\includegraphics[width=4.2cm]{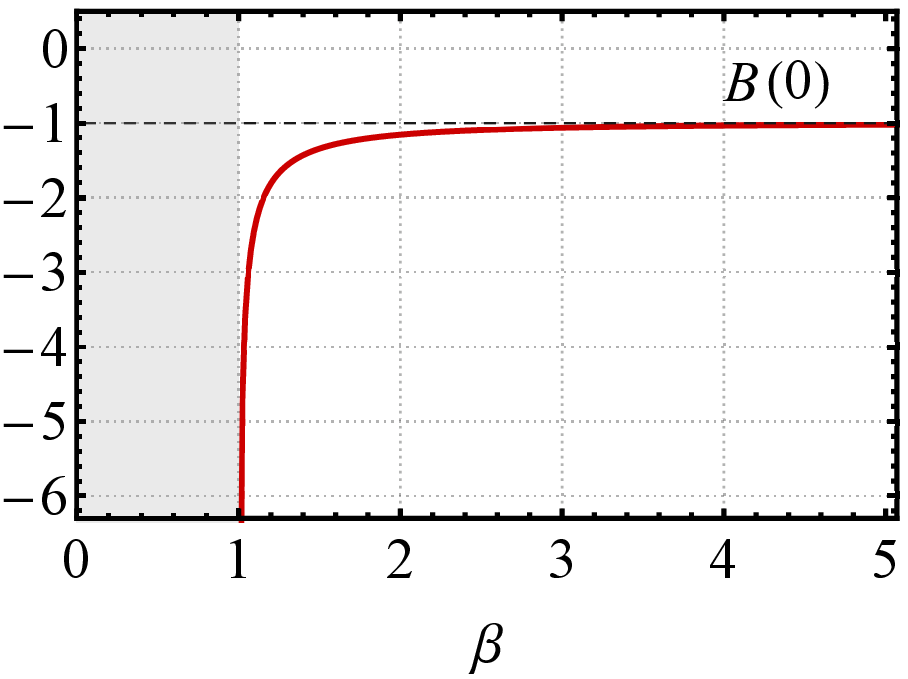}\hspace{0.1cm} %
	\includegraphics[width=4.12cm]{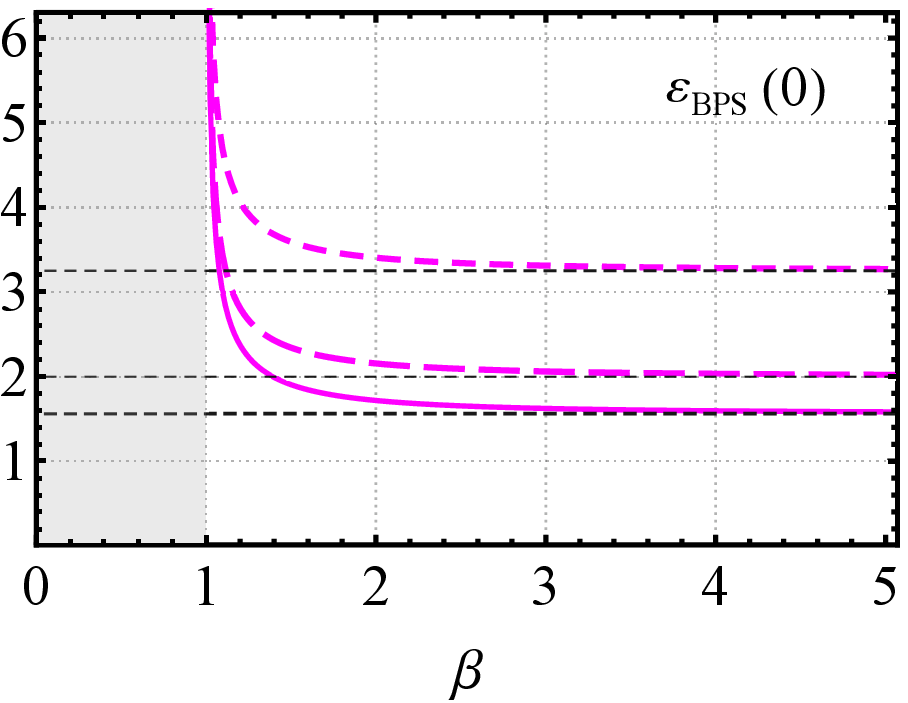}
	\caption{The magnetic field (left) and BPS energy density (right) amplitudes at the origin as functions of the BI parameter $\beta$, with $N=1$, $g=1$, and $\lambda=1$. The pictures depict both behaviors for the solitons coming from the superpotentials with $\sigma=3/2$ (solid line), $\sigma=2$ (long-dashed line), and $\sigma=3$ (dashed line), with the respective standard counterpart (black dashed lines).}
	\label{Fig07}
\end{figure}

\section{Conclusions and remarks \label{conclusion}}

{We have shown that topological BPS Skyrmions exist in a restricted baby Skyrme model endowed with a generalized gauge field dynamic given by the Born-Infeld term. The term possesses a free parameter $\beta$,  which for sufficiently large values reproduce effects similar to the  Maxwell term. Next, the Bogomol'nyi framework's successful implementation provides both an energy lower bound and the corresponding self-dual or BPS equations, whose solutions are the fields saturating such a lower limit.  The behavior near to the origin of the field profiles reveals the existence of a critical value (\ref{betaP}) to the BI parameter, i.e., only above such a value, we obtain well-behaved solitons.  On the other hand, depending on the behavior of the superpotential at large values of $r$, the model supports three different types of soliton solutions.  Forthcoming,  by choosing specific superpotentials, we have performed the numerical solutions of the BPS equations, and finally, we have compared the resulting solitons with those of the standard counterpart.}

{We have depicted, in Figs.  \ref{Fig01}  and \ref{Fig02}, the Born-Infeld compactons for different values of $\beta$, given a fixed value of $g$.  The figures allow us to conclude that the soliton becomes more compacted whenever the $\beta$ values decrease. The left-picture in the Fig. \ref{Fig06} depicts such property by showing clearly the dependence of compacton radius $R$  on $\beta$.  The shrinking of the soliton's radius becomes bigger when  $\beta$  is near to the critical value $\beta_{c}=1$ (demarcated by gray region). On the other hand, the soliton becomes more enlarged when  $\beta$ increases continuously until the maximum size of the soliton approaches that of the standard case.}

{Besides the compacton solitons, the model has two other types of solutions we named extended or noncompact solitons,  which also present some general features already described for the compactons, e.g.,  $\beta$ controls how the solutions move away from or approach the respective standard counterparts. The right-side of the Fig.  \ref{Fig06}  shows the evolution of vacuum value of the gauge field profile of the three types Born-Infeld Skyrmions as $\beta$ grows. It is clear how the values shifting towards those of the respective gauged BPS baby Skyrme model (dashed black lines). In addition, this plot also shows how the magnetic flux  (\ref{FlxB}) in units of $2\pi N$  changes as a function of the $\beta$ values.}

The values of the magnetic field at origin are independent of the $\sigma$ parameter according to the Eq. (\ref{B0}). This is shown in the bottom left-picture of Fig. \ref{Fig05} for the solutions with $\sigma>2$, and the left-picture into Fig. \ref{Fig07} shows the $B(0)$ amplitudes for all the superpotentials that coincide on an single curve. On the other hand, the BPS energy density $\varepsilon _{_{\text{BPS}}}(0)$ possesses an explicit dependence on the $\sigma$ values as given by the first term of the Eq. (\ref{B1}), see the lower right-picture in  Fig. \ref{Fig05} and right-picture of Fig. \ref{Fig07}.

Finally, we are now studying the possible existence of charged BPS Skyrmions (maybe behaving as anions) in the presence of dielectric media or magnetic impurities. Advances in this direction we will report elsewhere.

\begin{acknowledgments}
This study was financed in part by the Coordena\c{c}\~ao de Aperfei\c{c}%
oamento de Pessoal de N\'{\i}vel Superior - Brasil (CAPES) - Finance Code
001. We thank also the Conselho Nacional de Desenvolvimento Cient{\'\i}fico
e Tecnol\'ogico (CNPq), and the Funda\c{c}\~ao de Amparo \`a Pesquisa e ao
Desenvolvimento Cient{\'\i}fico e Tecnol\'ogico do Maranh\~ao (FAPEMA)
(Brazilian Government agencies). In particular, ACS thanks the full support
from CAPES. RC acknowledges the support from the grants CNPq/306724/2019-7,
CNPq/423862/2018-9, FAPEMA/Universal-01131/17, and FAPEMA/Universal-00812/19.
\end{acknowledgments}

\end{document}